\documentclass[final,12pt]{elsarticle}
\usepackage{graphicx}
\usepackage{multirow}%
\usepackage{amsmath,amssymb,amsfonts}%
\usepackage{amsthm}%
\usepackage{mathrsfs}%
\usepackage[title]{appendix}%
\usepackage{xcolor}%
\usepackage{textcomp}%
\usepackage{manyfoot}%
\usepackage{booktabs}%
\usepackage{algorithm}%
\usepackage{algorithmicx}%
\usepackage{algpseudocode}%
\usepackage{listings}%
\usepackage{multirow}
\usepackage[final]{microtype}
\usepackage{xurl}
\usepackage{subfig}
\usepackage{hyperref}
\hypersetup{
    colorlinks=true,
    linkcolor=blue,
    filecolor=magenta,      
    urlcolor=cyan,
    citecolor=blue,
    }
\usepackage{parskip}
\usepackage{lscape}

\usepackage{amssymb}

\begin{document}

\begin{frontmatter}



\title{Effect of color reconnection and rope formation on strange particle production in p+p collisions at $\sqrt{s}=13$~TeV 
}


\author[inst1]{Hushnud Hushnud\footnote{hushnud.jahan@gmail.com and hushnud.hushnud@cern.ch}}
\affiliation[inst1]{organization={Department of Physics},
            addressline={Aligarh Muslim University}, 
            city={Aligarh},
            postcode={202002}, 
            state={Uttar Pradesh},
            country={India}}

\author[inst2]{Kalyan Dey\footnote{kalyn.dey@gmail.com (corresponding author)}}
\affiliation[inst2]{organization={Department of Physics},
            addressline={Bodoland University}, 
            city={Kokrajhar},
            postcode={783370}, 
            state={Assam},
            country={India}}

\begin{abstract}
Strange particles are produced only during high-energy collisions and carry important information regarding collision dynamics. Recent results by the ALICE Collaboration on strangeness enhancement in high-multiplicity p+p collisions have highlighted the importance of the rope hadronization mechanism in high-energy nucleon-nucleon collisions. With the help of the \texttt{PYTHIA8} model, we made an attempt to study the strange particle production in high-energy p+p collisions at the LHC energy in the light of different color reconnection models and rope hadronization mechanism. The effect of color reconnection ranges on different observables is also discussed. The integrated yield of strange hadrons and baryon-to-meson ratios as a function of charged-particle multiplicity in p+p collisions at $\sqrt{s}$ = 13 TeV are well described by the hadronization mechanism of color ropes together with the QCD-based color reconnection scheme. The increasing trend of the average transverse momentum, $\langle p_{\rm T}\rangle$, as a function of $\langle dN/d\eta \rangle_{|\eta| < 0.5}$ can be explained quantitatively by the MPI-based color reconnection mechanism with a reconnection range of RR = 3.6; on the other hand, it is underestimated by the rope hadronization model.
\end{abstract}

\begin{keyword}
p+p collisions; high-multiplicity; \texttt{PYTHIA8}; rope hadronization; color reconnection; reconnection range
\end{keyword}

\end{frontmatter}

\section{Introduction}
\label{sec1}
Observables measured in high-energy nucleus-nucleus (A+A) collisions suggested the formation of deconfined hot and dense matter known as Quark-Gluon Plasma (QGP) \cite{Busza:2018}. Collective flow is one of the characteristic features of this hot and dense medium of strongly interacting matter where the produced medium after achieving thermal equilibrium, expands collectively with an average flow velocity equal for each particle species. The collective flow phenomena are expected to produce a mass ordering of the average transverse momentum, $\langle \mathrm{p_T}\rangle$, for the produced particles, which in fact is seen in nucleus-nucleus collisions as reported by various heavy-ion collision experiments at the RHIC and the LHC \cite{Sahu_2006, Khachatryan:2017, Abelev_2013, Abelev_2014, Abelev_2015}. Of late, the heavy-ion physics community has been fascinated by the observation of unexpected collective behavior in high-multiplicity (HM) p+p collision events. The heavy-ion-like collective behavior is reflected in the ridge-like structure on the near side of two-particle correlations observed by the CMS, the ATLAS, and the ALICE Collaboration at the LHC \cite{Khachatryan:2010, Khachatryan:2016, Velicanu:2011}. This ridge-like structure is believed to be caused by the long-range correlations in particle production. The heavy-ion-like effect is also seen in the strangeness sector. The enhancement of strange hadrons in nucleus-nucleus collisions with respect to baseline p+p collisions is known to be one of the traditional signatures of the formation of the QGP \cite{Rafelski:1986, Koch:1986}. Enhanced production of strange hadrons relative to pions in HM p+p collision events have been reported by the ALICE Collaboration \cite{nature:2017}. The magnitude of the enhancement is found to be similar to those observed in Pb+Pb collisions where a QGP is produced. Furthermore, a strange-quark ordering in strangeness enhancement is observed in HM p+p events. The above observation is very similar to the one seen in the heavy-ion collision data reported by various experiments at the SPS, the RHIC, and the LHC \cite{Andersen_1998, Andersen_1999, Afanasiev_2002, Antinori_2004, Anticic_2004, Adams_2004, Collaboration_2006, Abelev_2008, MultAbelev_2013, MultAbelev_2014}. Currently, it is quite unclear if the creation of QGP droplets or any other non-QGP effects in small systems is responsible for such heavy-ion-like behavior. 
Monte Carlo (MC) models are often utilized as a tool to explain experimental findings in the context of underlying physics processes. 
Among these models, \texttt{PYTHIA} stands out as one of the most extensively employed and successful tools for interpreting diverse experimental observations in pp collisions. Section~\ref{sec:2} provides a concise overview of the \texttt{PYTHIA} model. It is worth emphasizing that the implementation of the Monash tune in \texttt{PYTHIA} has proven to be effective in addressing certain shortcomings of the earlier versions of \texttt{PYTHIA} especially in the soft QCD regime. For a detailed description of the Monash settings, the reader may go through the Ref.~\cite{Monash}. The Monash tune of \texttt{PYTHIA} involves the adjustment of various parameters related to different physics processes, encompassing initial state radiation (ISR), final state radiation (FSR), string breaking phenomena, parton distribution function, matrix element, Multi-Parton-Interaction (MPI), Colour-Reconnection (CR), etc. The fine-tuned \texttt{PYTHIA} (Monash) is found to explain various experimental findings including the centrality-dependent average transverse momentum, $\langle p_{\rm T} \rangle$, for charged particles at the LHC energies \cite{ALICE:2017, ALICEEPJC:2017, ALICEPLB:2013, Acharya_2019}. Despite achieving numerous successes, it is important to highlight that \texttt{PYTHIA} (Monash) falls short in explaining the strange baryon multiplicity and the observed strangeness enhancement in pp collisions at LHC energies. However, the incorporation of the color rope formation mechanism in the \texttt{PYTHIA} model to some extent is able to explain the ratio of yield of strange baryons with respect to charged pions in pp collisions at the LHC energies \cite{Bierlich_2019, Nayak_2019, Hussein:2022, loizides2021apparent, Cui_2022}. Another MC model called DIPSY rope \cite{Flensburg_2011}, a QCD-inspired model based on Mueller's dipole formulation, is seen to be quite successful in the strangeness sector explaining the production of single strange hadrons (K$^{0}_{s}$ and $\Lambda$) while it tends to underestimate the production of multi-strange baryons ($\Xi$ and $\Omega$). In the DIPSY rope model, color ropes emerge from the interaction of gluonic strings, leading to heightened string tension and, consequently, increased production of strange particles. EPOS-LHC \cite{Werner:2012} is considered as one of the most effective Monte Carlo models for simulating proton-proton collisions, even though it falls short in accounting for the observed strangeness enhancement at the LHC energies. In the EPOS model, the Gribov-Regge theory is applied to describe the dynamics of hard scattering in the ``core" region where the string density is large. While for the less dense region surrounding the core called the corona, the EPOS model incorporates non-perturbative aspects of QCD, involving multiple parton interactions, gluon exchanges, and other softer processes that contribute to the overall hadronization and final-state particle production.
Even though the three Monte Carlo models (\texttt{PYTHIA}-Monash, DIPSY rope, EPOS) qualitatively describe the evolution of mean transverse momentum, $\langle p_{\rm T} \rangle$ with multiplicity at LHC energies, there exists a disparity in quantitative concordance with experimental data. In terms of quantitative assessment, EPOS-LHC offers a marginally improved portrayal of the multiplicity evolution of $\langle p_{\rm T} \rangle$ for strange baryons \cite{Acharya_pp:2020}. In a recent study \texttt{PYTHIA} model with different magnitude of color reconnection range (RR) was used to investigate the multiplicity evolution of mean transverse momentum, $\langle p_{\rm T} \rangle$ of the identified particles \cite{meanpt1}. It was observed that a reconnection range of RR = 3.6 provides a better representation of the experimental data for kaons and protons in p+p collisions at $\sqrt{s}$ = 13 TeV. In this paper, an attempt has therefore been made to study the effect of color reconnection ranges on the strange particle dynamics, especially the evolution of $\langle p_{\rm T} \rangle$ with multiplicity in p+p collisions at $\sqrt{s}$ = 13 TeV. The effect of various CR schemes and color rope formation mechanisms on strange (K$^{0}_{s}$ and $\Lambda$) and multi-strange hadrons ($\Xi$ and $\Omega$) production in PYTHIA8 is also discussed and compared with the available experimental data. 

The paper is structured as follows. A detailed description of the \texttt{PYTHIA8} event generator is discussed in Section \ref{sec:2}. In Section \ref{sec:3}, the event generation and analysis methodologies are described. The multiplicity-dependent integrated yield $\langle dN/dy\rangle$, average transverse momentum $\langle p_{\rm T} \rangle$, and the baryon-to-meson ratios for strange hadrons are presented in Section \ref{sec:4}. Finally, in Section \ref{sec:5}, we will summarize the important outcomes of the present work.  

\section{The \texttt{PYTHIA8} Model}\label{sec:2}
\texttt{PYTHIA8} is a widely used Monte Carlo event generator generally used for simulating p+p, $e^{+}$+$e^{-}$, and $\mu^{+}$+$\mu^{-}$ collisions at relativistic energies. It is a pQCD-based model that incorporates factorized perturbative expansion for the hard parton-parton scattering, the initial and final state parton showers, various models for hadronization, and multiparton interactions \cite{PYTHIA1}.
In \texttt{PYTHIA8}, the hadronization process is modeled via the Lund string model \cite{PYTHIA2, PYTHIA3}. In this model, the massless string is assumed to be formed due to the color field existing between two interacting partons leading to linear confinement. As the string potential energy increases, the partons start to move apart, and hence the string breaks, causing the production of lighter parton pairs. These partons then undergo interactions in the overlap region in hadronic collisions. These constitute the MPI framework in the \texttt{PYTHIA8} model. One of the most important features incorporated in \texttt{PYTHIA8} is the CR mechanism, where the strings are rearranged between partons in such a way that the total string length is minimized. This mechanism is capable of explaining a variety of heavy-ion-like effects observed in p+p collisions \cite{PhysRevLett:ortiz, PhysRevD:Christian, Dey:2022}. Currently, three variants of CR mechanisms are incorporated in the \texttt{PYTHIA8} model. The default variant, i.e., the MPI-based mechanism \cite{PYTHIA4, PYTHIA5} is based on the phenomenon in which the partons of a lower-$p_{\rm T}$ MPI system connect with one in a higher-$p_{\rm T}$ MPI system to reduce the total string length. The next variant, known as the QCD-based CR model \cite{PYTHIA4}, is the one where the reconnection is allowed to happen based on QCD color rules if and only if there is a reduction in the total string length and the string potential energy. The third variant of CR is based on the gluon move-based mechanism \cite{christiansen2015colour, PYTHIA1} in which the partons can be moved from one location to another to reduce the total string length. The rope hadronization model of \texttt{PYTHIA8} is an extension of the Lund string hadronization model. This model suggests that several strings that are very close to each other can fuse to form colored ropes. These color ropes have a larger effective string tension, and hence they fragment into more strange quarks and diquarks. This fragmentation leads to enhanced production of baryons and strange hadrons \cite{PYTHIA7, PYTHIA8}. 
The string interaction in rope hadronization is described by two mechanisms: string-shoving and flavor ropes. The string-shoving model \cite{PYTHIA1} allows the nearby strings to shove each other with an interaction potential derived from the color superconductor analogy. However, the flavor ropes allow the formation of ropes between strings overlapping in a dense environment, which is hadronized with a larger effective string tension.     

\section{Event generation and analysis details} \label{sec:3}
\texttt{PYTHIA 8.306} is used to generate 50 million inelastic non-diffractive p+p collisions at $\sqrt{s}= 13$~TeV in each configuration. In the present analysis, we have used the \texttt{PYTHIA8} Monash 2013 tune \cite{Monash} with multiparton interactions (MPI) and color reconnection mechanisms \cite{PYTHIA4, PYTHIA5, christiansen2015colour, PYTHIA1} (MPI-based CR and QCD-based CR) and rope hadronization (RH) \cite{PYTHIA7, PYTHIA8, PYTHIA1}. All the resonances are allowed to decay except for the ones used in the current analysis. In this analysis, only those events were selected for which at least one charged-particle is produced in the mid-rapidity range $|\eta|<1$.
For the multiplicity estimation, charged particles are measured in the acceptance of the V0 detectors of the ALICE experiment, which has pseudo-rapidity coverage of V0A (2.8 $<$ $\eta$ $<$ 5.1) and V0C (-3.7 $<$ $\eta$ $<$ -1.7). 
These events are further divided into ten V0 multiplicity (V0M) bins. However, the mean charged-particle density, $\langle dN_{ch}/d\eta\rangle$ has been estimated at mid-rapidity ($|\eta| < 0.5$). The multiplicity classes and the corresponding charged-particle multiplicities used by the ALICE experiment and the different mechanisms of \texttt{PYTHIA8} are depicted in Table~\ref{centbin}.  
It is noted that the mean charged-particle density, $\langle dN_{ch}/d\eta \rangle_{|\eta| < 0.5}$, of primary charged particles is quite large when the CR mechanism is not incorporated. 
This observation can be explained by the fact that in the absence of the CR mechanism, the individual particle interactions are independent of each other \cite{MeanMult}, which might result in enhanced production of charged particles.

\section{Results and Discussion}\label{sec:4}
In the present analysis, the study of strange hadrons (K$^{0}_{s}$, $\Lambda$, $\Xi$, and $\Omega$) has been performed in the mid-rapidity window ($|y|$ $<$ 0.5). To understand the production mechanism of strange hadrons in p+p collisions, various observables, such as the $p_{\rm T}$-integrated yield, average transverse momentum $\langle p_{\rm T}\rangle$, particle ratio, etc., as a function of $\langle dN_{ch}/d\eta \rangle_{|\eta| < 0.5}$, are calculated by employing different configurations of the \texttt{PYTHIA8} model and compared with the available data from the ALICE experiment. 

\subsection{Integrated yields }\label{Integrated_yield}
\begin{figure}[h]
\centerline{\includegraphics[width=5.3in]{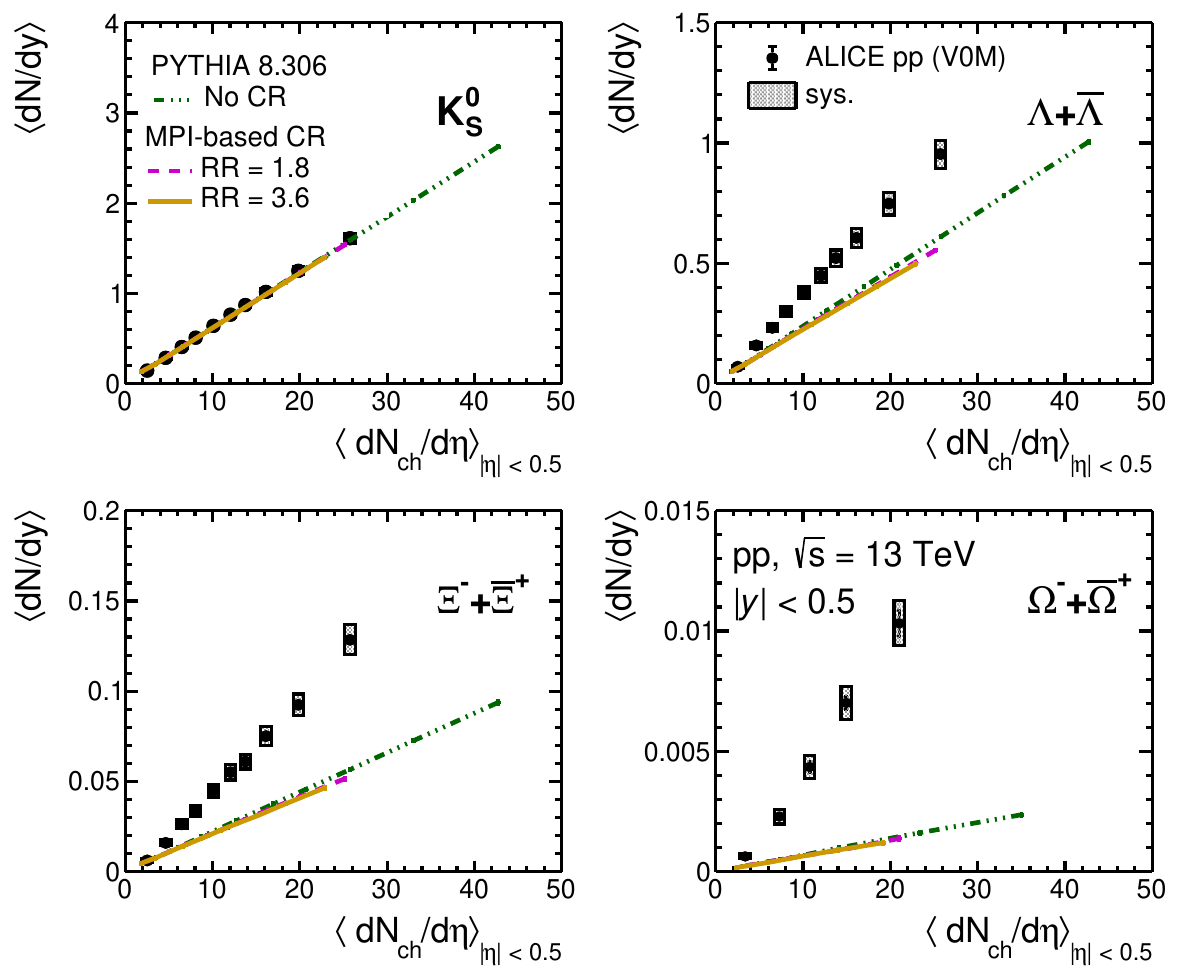}}
\caption{Integrated yield of K$^0_s$ meson, $\Lambda$, $\Xi$ and $\Omega$ baryons at mid-rapidity as a function of $\langle dN_{ch}/d\eta \rangle_{|\eta| < 0.5}$ in p+p collisions at $\sqrt{s}=13$ TeV using \texttt{PYTHIA8} with MPI-based CR mode for various RR. The results are compared with the integrated yield measured by the ALICE experiment \cite{multpp13}.\protect\label{fig:dndy_Mode0}}
\end{figure}

Integrated yields, $\langle dN/dy \rangle$, of strange meson (K$^{0}_{s}$) and strange baryons ($\Lambda$, $\Xi$, and $\Omega$) at mid-rapidity are plotted as a function of $\langle dN_{ch}/d\eta \rangle_{|\eta| < 0.5}$ as shown in Fig.~\ref{fig:dndy_Mode0} for p+p collisions at $\sqrt{s}$ = 13 TeV  using the \texttt{PYTHIA8} model. These results are compared with the available experimental data from the ALICE experiment \cite{multpp13}. Fig.~\ref{fig:dndy_Mode0} displays the results from the \texttt{PYTHIA8} model with MPI-based CR mechanism for two different color reconnection ranges, RR = 1.8 and 3.6. For comparison, we have also shown the results for the events without the CR mechanism. It can be clearly seen from Fig.~\ref{fig:dndy_Mode0} that, except for the K$^{0}_{s}$ meson, the MPI-based CR mechanism underestimates the experimental data. The discrepancy increases with increasing multiplicity as well as with the increase in strange quark content. 
The parameters of the MPI-based CR in \texttt{PYTHIA} model has been tuned to reproduce experimental data on non-strange hadrons. Although the presence of MPI introduces additional partonic interactions that can affect the color structure of the final state hadrons through color reconnection, its implementation in \texttt{PYTHIA} might not explicitly account for the production of strange quarks during these additional soft interactions\footnote{It is to be noted here that the strangeness suppression factor in the LUND string fragmentation used in \texttt{PYTHIA} i.e. \texttt{StringFlav:probStoUD} is 0.217 (default)}.
Thus, the inadequate modeling of soft or non-perturbative processes could result in disparities between predictions and experimental data concerning multi-strange baryons.
It can also be seen from Fig.~\ref{fig:dndy_Mode0} that there is no significant effect of reconnection radius (RR) on the integrated yield of strange hadrons.

\begin{landscape}
\begin{table}[h]
\caption{\large{Various V0M multiplicity classes and the corresponding average charged-particle density ($\langle dN_{ch}/d\eta \rangle_{|\eta| < 0.5}$) in p+p collisions at $\sqrt{s}$ = 13 TeV used by the ALICE experiment \cite{multpp13} and different \texttt{PYTHIA8} models.}}
\vspace{1em}
{\normalsize
\begin{tabular}{cccccccc}
\toprule
\multirow{4}{2em}{Event \centering Classes} & \multirow{4}{4em}{Centrality \centering ($\%$)} & \multicolumn{5}{c}{$\langle dN_{ch}/d\eta \rangle$} \\
\cmidrule{3-8}
& & \multirow{3}{6em}{Experimental \centering Data} & \multicolumn{4}{c}{PYTHIA8 Model} \\
\cmidrule{4-8} 
& & & \multirow{2}{*}{No CR} & \multicolumn{2}{c}{MPI-based CR Scheme} & \multirow{2}{3cm}{QCD-based CR \centering Scheme} &  \multirow{2}{*}{RH} \\
\cmidrule{5-6}   
& & & & RR = 1.8 & RR = 3.6 & & \\
\midrule
I & 0.0-0.9 & 25.75 $\pm$ 0.40 & 42.713 $\pm$ 0.017 & 25.167 $\pm$ 0.014 & 22.859 $\pm$ 0.013 & 24.927 $\pm$ 0.014 &23.846 $\pm$ 0.013\\

II & 0.9-4.5 & 19.83 $\pm$ 0.30 & 33.011 $\pm$ 0.008 & 20.056 $\pm$ 0.006 & 18.381 $\pm$ 0.005 & 20.134 $\pm$ 0.006 &19.341 $\pm$ 0.006\\

III & 4.5-8.9 & 16.12 $\pm$ 0.24 & 25.798 $\pm$ 0.006 & 16.378 $\pm$ 0.005 & 15.107 $\pm$ 0.004 & 16.479 $\pm$ 0.005 &15.830 $\pm$ 0.005\\

IV & 8.9-13.5 & 13.76 $\pm$ 0.21 & 20.708 $\pm$ 0.005 & 13.780 $\pm$ 0.004 & 12.797 $\pm$ 0.004 & 13.828 $\pm$ 0.004 &13.200 $\pm$ 0.004\\

V & 13.5-18.0 & 12.06 $\pm$ 0.18 & 16.977 $\pm$ 0.005 & 11.770 $\pm$ 0.004 & 10.934 $\pm$ 0.004 &11.774 $\pm$ 0.00 & 11.310 $\pm$ 0.004\\

VI & 18.0-27.0 & 10.11 $\pm$ 0.15 & 12.712 $\pm$ 0.003 & 9.390 $\pm$ 0.002 & 8.722$\pm$ 0.002 & 9.402 $\pm$ 0.002 & 9.199 $\pm$ 0.002\\

VII & 27.0-36.1 & 8.07 $\pm$ 0.12 & 8.678 $\pm$ 0.002 & 6.966 $\pm$ 0.002 & 6.597 $\pm$ 0.002 & 7.016 $\pm$ 0.002 & 7.049 $\pm$ 0.002\\

VIII & 36.1-45.3 & 6.48 $\pm$ 0.10 & 6.034 $\pm$ 0.002 & 5.182 $\pm$ 0.002 & 5.056 $\pm$ 0.002 & 5.271 $\pm$ 0.002 & 5.530 $\pm$ 0.002\\

IX & 45.3-64.5 & 4.64 $\pm$ 0.07 & 3.583 $\pm$ 0.001 & 3.366 $\pm$ 0.001 & 3.379 $\pm$ 0.001 & 3.462 $\pm$ 0.001 & 3.921 $\pm$ 0.001\\

X & 64.5-100.0 & 2.52 $\pm$ 0.04 & 1.799 $\pm$ 0.001 & 1.852 $\pm$ 0.001 & 1.862 $\pm$ 0.001 & 1.793 $\pm$ 0.001 & 1.956 $\pm$ 0.001\\
\bottomrule
\end{tabular}\label{centbin}}
\end{table}
\end{landscape}

\begin{figure}[pht]
\centerline{\includegraphics[width=5.3in]{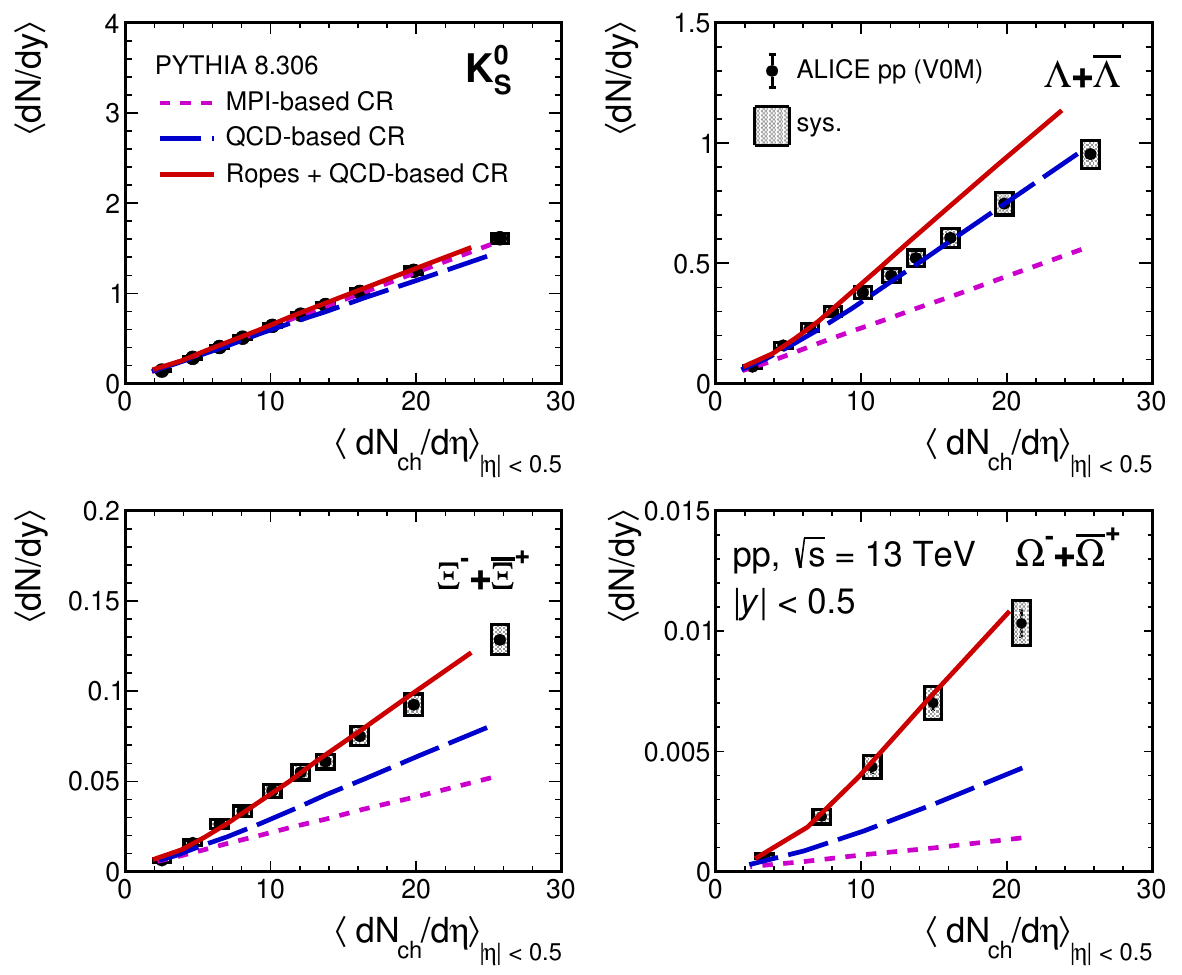}}
\caption{A multiplicity-dependent $\langle dN/dy \rangle$ of K$^0_s$ meson, $\Lambda$, $\Xi$ and $\Omega$ baryons at mid-rapidity in p+p collisions at $\sqrt{s}=13$ TeV using \texttt{PYTHIA8} MPI-based CR, QCD-based CR and RH for RR = 1.8. The results are compared with the integrated yield measured by the ALICE experiment \cite{multpp13}.\protect\label{fig:dndy_RR18}}
\end{figure}

In Fig.~\ref{fig:dndy_RR18}, multiplicity-dependent $\langle dN/dy \rangle$ for strange hadrons is shown using \texttt{PYTHIA8} with MPI-based CR, QCD-based CR mechanism, and rope hadronization (RH). It is seen from Fig.~\ref{fig:dndy_RR18} that the MPI-based CR describes the $\langle dN/dy \rangle$ of K$^{0}_{s}$ meson, whereas it underestimates the production of strange baryons ($\Lambda$, $\Xi$, and $\Omega$). However, the QCD-based CR scheme, which enables the formation of junctions formed by joining the three color lines, leading to enhanced production of baryons, explains the yields of $\Lambda$ containing a single strange quark. 
This investigation also reveals that the QCD-based CR is insufficient to fully explain the production of multi-strange baryons. Consequently, there is a need for additional mechanisms related to strangeness production to provide a comprehensive account of the experimental data.
Moreover, the yield of K$^{0}_{s}$ mesons at higher multiplicities is found to be marginally reduced in the QCD-based Color Reconnection (CR) model compared to the MPI-based CR model. This finding is contrary to the trend that is observed for baryons. The higher probability of baryon production in the QCD-based CR model is generally associated with the lowering of meson production. 
On the other hand, the rope hadronization phenomena in combination with the QCD-based CR mechanism further increases the effective string tension and thus increases the overall production of baryons. The yield of $\Lambda$ baryons which was already explained by the QCD-based CR model is therefore overestimated by the rope hadronization (RH) mechanism. The RH in combination with the QCD-based CR mechanism is found to explain the experimental data for multi-strange baryons such as $\Xi$ and $\Omega$.

\subsection{Average transverse momentum}

\begin{figure}[pht]
\centerline{\includegraphics[width=5.3in]{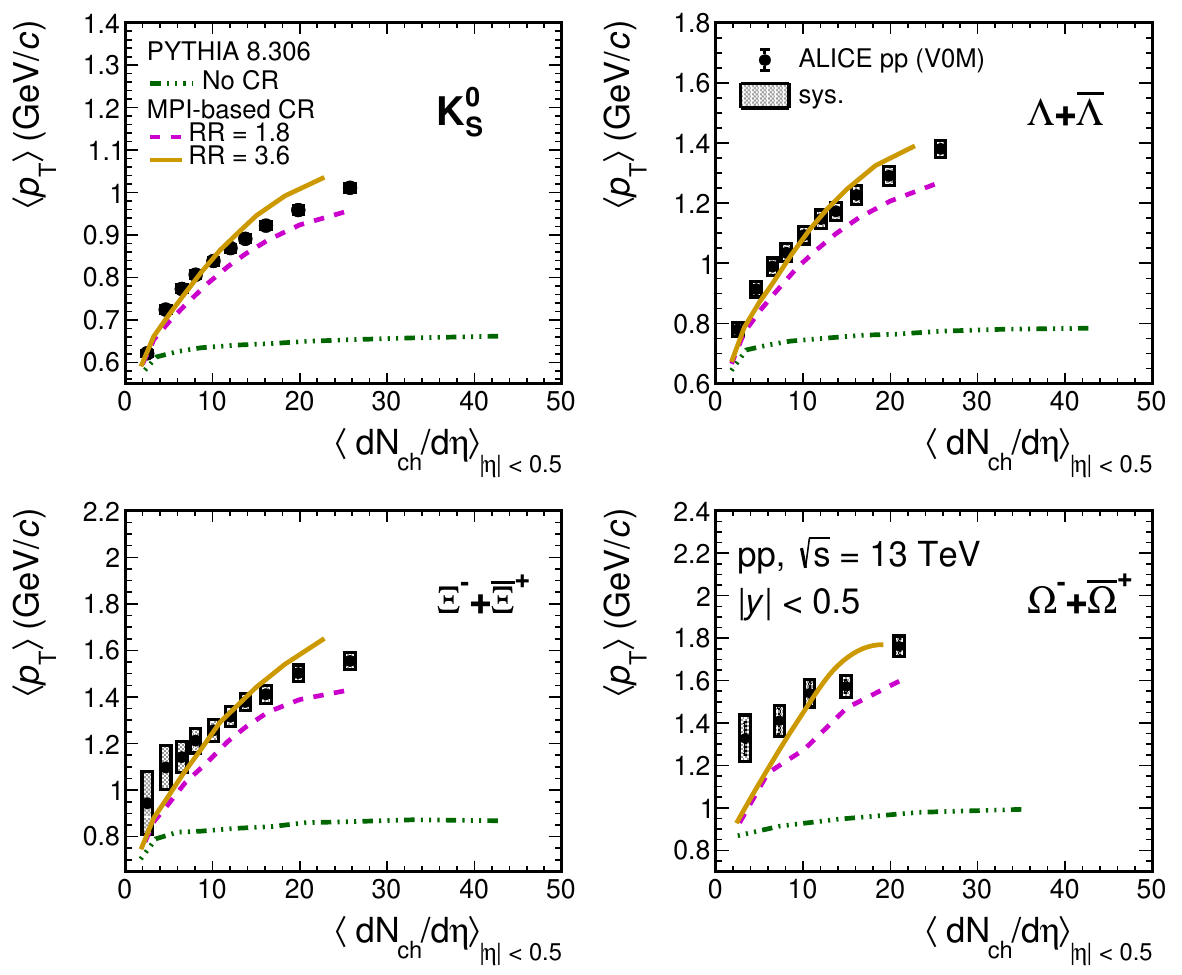}}
\caption{Average transverse momentum, $\langle p_{\rm T} \rangle$, for K$^0_s$ meson, $\Lambda$, $\Xi$, and $\Omega$ baryons at mid-rapidity as a function of $\langle dN_{ch}/d\eta \rangle_{|\eta| < 0.5}$ in p+p collisions at $\sqrt{s}=13$ TeV using \texttt{PYTHIA8} MPI-based CR mode for various RR and are compared with the results of the ALICE experiment \cite{multpp13}.\protect\label{fig:meanpt_Mode0}}
\end{figure}

\begin{figure}[pht]
\centerline{\includegraphics[width=5.3in]{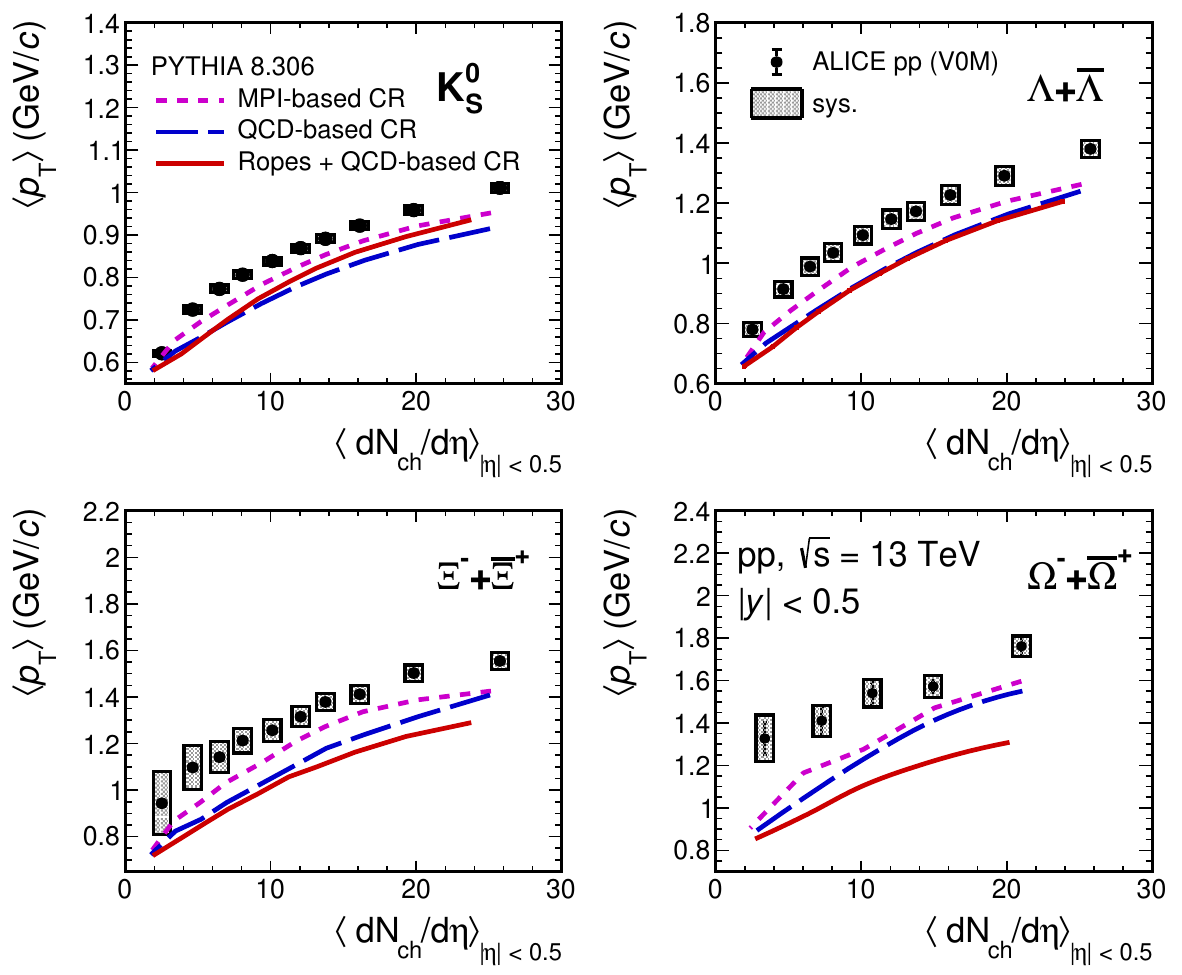}}
\caption{Multiplicity-dependent average transverse momentum, $\langle p_{\rm T} \rangle$, for K$^0_s$ meson, $\Lambda$, $\Xi$ and $\Omega$ baryons at mid-rapidity as a function of $\langle dN_{ch}/d\eta \rangle_{|\eta| < 0.5}$ in p+p collisions at $\sqrt{s}=13$ TeV using \texttt{PYTHIA8} MPI-based CR, QCD-based CR and RH for RR = 1.8 and are compared with the results of the ALICE experiment \cite{multpp13}.\protect\label{fig:meanpt_RR18}}
\end{figure}

To shed some light on the onset of collective behavior of the strange hadrons produced in the small system, a study has been performed where the multiplicity-dependent average transverse momentum, $\langle p_{\rm T} \rangle$, of strange hadrons (K$^{0}_{s}$, $\Lambda$, $\Xi$, and $\Omega$) has been calculated. Figs.~\ref{fig:meanpt_Mode0} and ~\ref{fig:meanpt_RR18} illustrate  the $\langle p_{\rm T} \rangle$ of the studied strange particles at mid-rapidity as a function of $\langle dN_{ch}/d\eta \rangle_{|\eta| < 0.5}$ in p+p collisions at $\sqrt{s}$ = 13 TeV for different variants of the \texttt{PYTHIA8} model and compared with the available experimental data from the ALICE experiment \cite{multpp13}. The increasing trend of $\langle p_{\rm T} \rangle$ with multiplicity, as observed in both \texttt{PYTHIA8} and the ALICE data \cite{multpp13}, is believed to be an outcome of the hardening of $p_{\rm T}$-spectra with increasing multiplicity. It is important to note that the increase in the value of $\langle p_{\rm T}\rangle$ as a function of charged-particle multiplicity is more pronounced in the case of heavier strange particles. This mass-dependent hardening observed in \texttt{PYTHIA8} is a consequence of the CR mechanism  as already reported in \cite{meanpt1, PhysRevLett:ortiz, Adam_2016, Ortiz_2017, Ortiz_Velasquez_2019} and can also be argued from Fig.~\ref{fig:meanpt_Mode0} where \texttt{PYTHIA8} without the CR mechanism is not at all able to explain the measured $\langle p_{\rm T} \rangle$ by the ALICE experiment for strange hadrons. 
It is to be noted here that while \texttt{PYTHIA8} with the MPI-based CR model (RR=1.8) successfully describes the observed trend in the average transverse momentum, $\langle p_{\rm T}\rangle$, as a function of $\langle dN_{ch}/d\eta \rangle_{|\eta| < 0.5}$  measured by the ALICE experiment, there is still a noticeable absence of exact quantitative agreement between the model and experimental results.
The discrepancy increases as one moves toward the hadrons with higher strangeness content. One can see from Fig.~\ref{fig:meanpt_Mode0} that the MPI-based CR model with RR = 3.6 can describe the experimental data for almost all multiplicity bins. However, the model slightly overestimates $\langle p_{\rm T} \rangle$ of K$^{0}_{s}$ meson and $\Xi$ baryon towards higher multiplicity, whereas it underestimates the $\langle p_{\rm T} \rangle$ of $\Omega$ baryon for the lowest multiplicity bin. A slight hint of discontinuity at $\langle dN_{ch}/d\eta \rangle_{|\eta| < 0.5} \sim 15$  seen in Fig.~\ref{fig:meanpt_Mode0} for $\Omega$ baryon with RR=3.6 can be attributed to the statistical fluctuation associated with the highest multiplicity bin.

In Fig.~\ref{fig:meanpt_RR18}, the multiplicity-dependent $\langle p_{\rm T} \rangle$ of strange hadrons from the MPI-based CR, QCD-based CR and RH models are depicted and compared with the ALICE data \cite{multpp13}. 
Although all the variants of the \texttt{PYTHIA8} model were found to underestimate the experimental data, the MPI-based CR tune provided a comparatively better description of the data. 
It can further be seen from the figure that the deviation from experimental data increases with the increasing strange quark content of hadrons. 
It is interesting to note that the RH model that explains the integrated yields of strange hadrons totally fails to explain the multiplicity-dependent $\langle p_{\rm T} \rangle$. 
It is to be noted here that the inclusion of rope formation mechanism to the existing QCD-based CR model leads to a higher number of strange quarks and di-quarks being produced. As a result, the average transverse momentum $\langle p_{\rm T} \rangle$ of these particles created decreases. This happens because more energy is used in particle creation, leaving less energy available for ``transverse boost", which pushes particles sideways. Since strange quarks are heavier than light quarks, they require more energy for the same level of transverse boost. This effect is more pronounced for $\Xi$ and $\Omega$ baryons, containing two and three strange quarks, respectively. For $\Lambda$ baryons, containing only one strange quark, this effect is less noticeable. Thus, the QCD-based CR and Rope + QCD-based CR predictions are consistent with each other as far as the multiplicity dependent $\langle p_{\rm T} \rangle$ is concerned for $\Lambda$ baryons.

\subsection{Particle ratios}

\begin{figure}[pht]
\centerline{
\hspace*{-5em}
\includegraphics[width=6in]{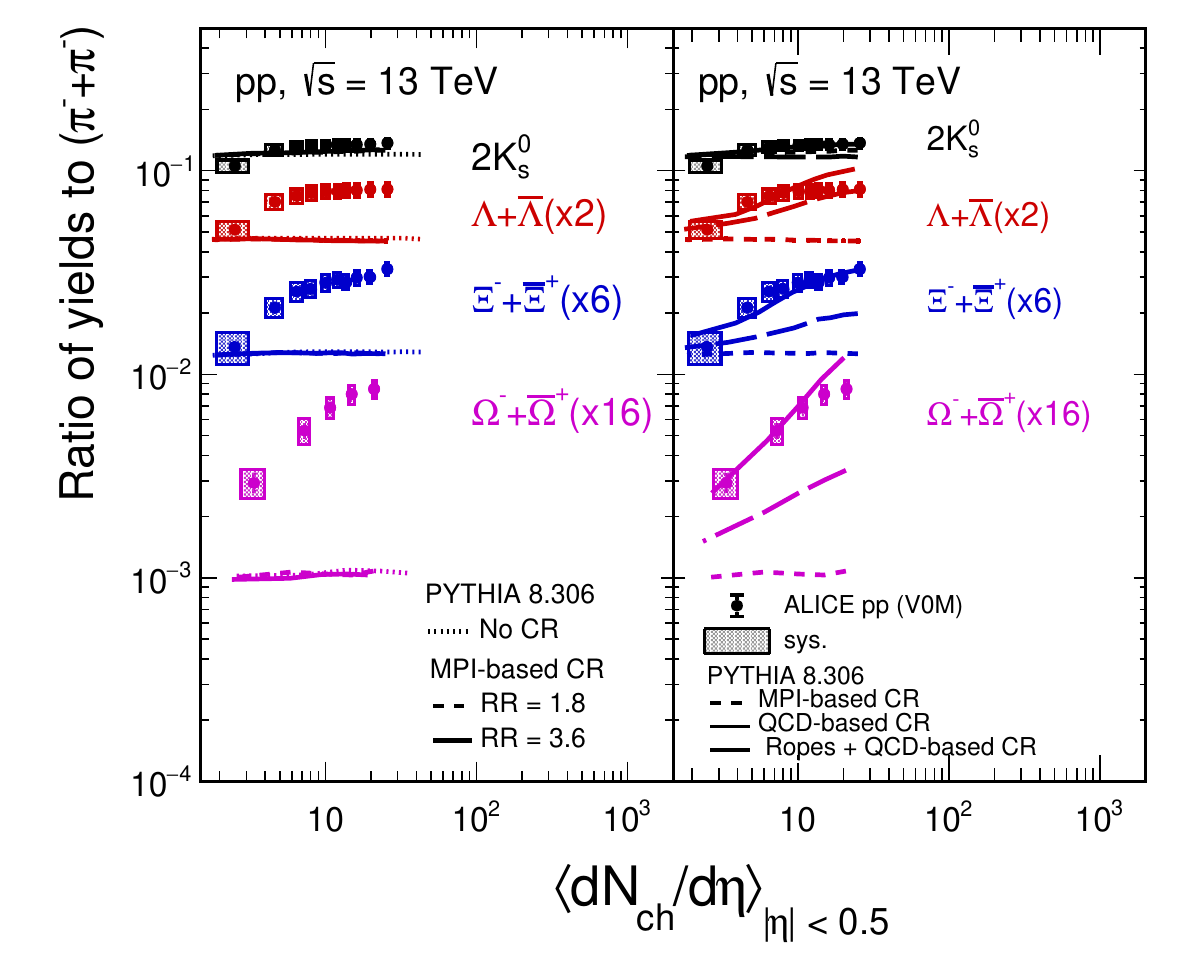}}
\caption{\textit{Left:} The ratio of yields of hadron (K$^{0}_{s}$, $\Lambda$, $\Xi$ and $\Omega$)-to-pion ($\pi^{+} + \pi^{-}$) at mid-rapidity as a function of $\langle dN_{ch}/d\eta \rangle_{|\eta| < 0.5}$ in p+p collisions at $\sqrt{s}=13$ TeV using different RR of the \texttt{PYTHIA8} model. \textit{Right:} The same is plotted for different variants of \texttt{PYTHIA8} model. The results are compared with the measured particle ratios from the high-multiplicity (HM) p+p collision events high-multiplicity (HM) p+p collision events ALICE experiment \cite{multpp13}.
\protect\label{fig:BaryonbypionMode0}}
\end{figure}

To study the enhanced production of strange hadrons in small systems as reported by the ALICE experiment \cite{nature:2017}, the ratio of yields of strange hadrons (K$^{0}_{s}$, $\Lambda$, $\Xi$, and $\Omega$) with respect to non-strange particles ($\pi^{+}$+$\pi^{-}$) within the same acceptance has been estimated. In Fig.~\ref{fig:BaryonbypionMode0} (left), these ratios are plotted as a function of $\langle dN_{ch}/d\eta  \rangle_{|\eta| < 0.5}$ by using the \texttt{PYTHIA8} model with different values of RR and compared with the results from the ALICE experiment \cite{PartRatio1}. It can clearly be seen from this figure that the \texttt{PYTHIA8} model with MPI-based CR mechanism underestimates the ALICE data except for K$^{0}_{s}$/$\pi$ ratio. As expected, the RR plays no significant role in explaining the particle ratio which can also be evident from Fig.~\ref{fig:dndy_Mode0}. The discrepancy between the experimental data and the model calculations widens with the increase of strange quark content of hadrons. The rationale behind the reduced yield of strange baryons in MPI-based CR is explained in Section \ref{Integrated_yield}. 

\begin{figure}[pht]
\centerline{
\hspace*{-5em}
\includegraphics[width=6in]{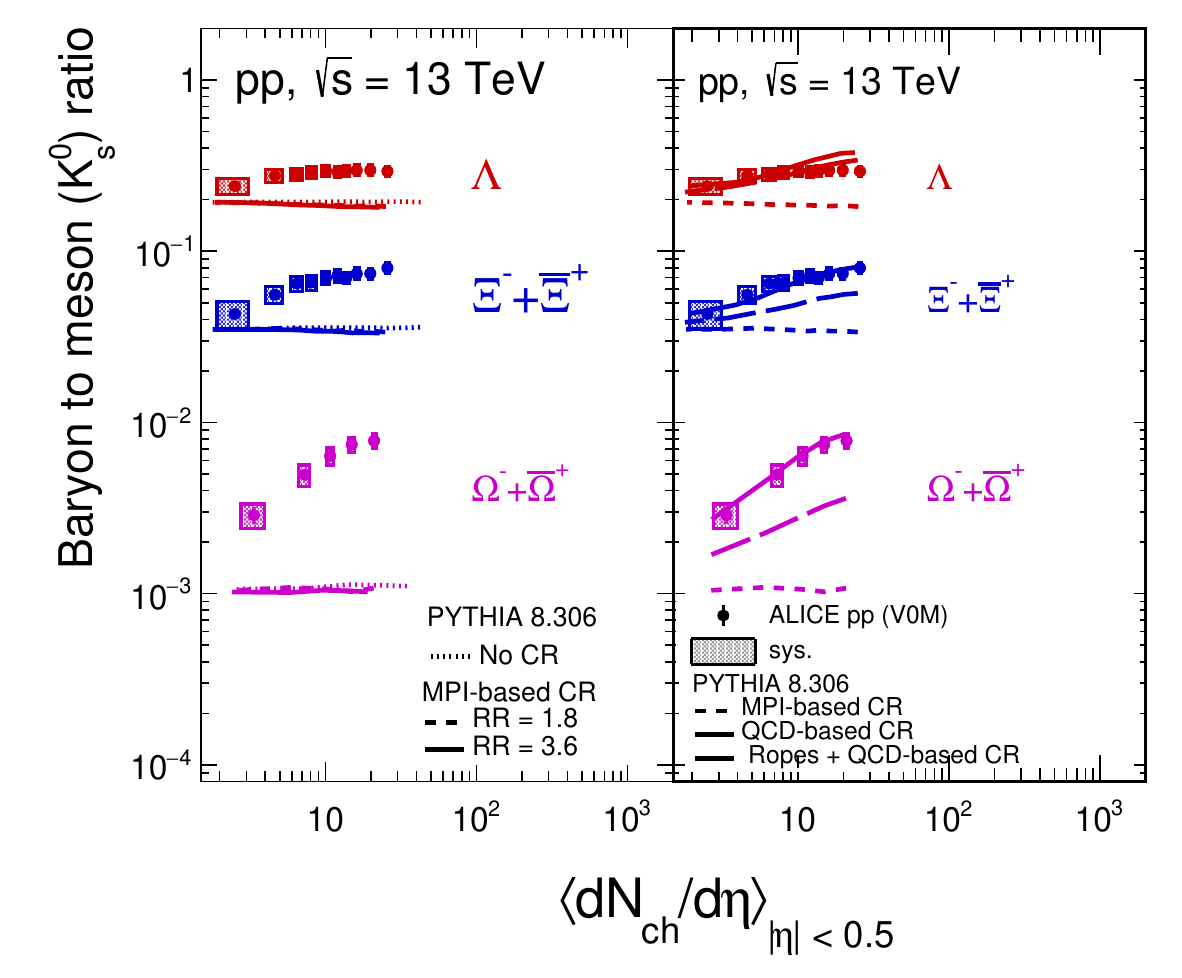}}
\caption{The multiplicity-dependent baryon ($\Lambda$, $\Xi$, and $\Omega$)-to-meson(K$^{0}_{s}$) ratios at mid-rapidity in p+p collisions at $\sqrt{s}=13$ TeV using different values of RR for MPI-based CR mechanism in the \texttt{PYTHIA8} model (Left) and for different schemes of \texttt{PYTHIA8} (Right) and compared with the ALICE data \cite{multpp13}. It is to be noted that for this particular case, $\Lambda$ contains no contribution from $\bar{\Lambda}$ baryons.\protect\label{fig:BaryonbyK0s_Mode0}}
\end{figure}

Further, to gain more insight into the production of strange particles, the multiplicity-dependent strange-to-non-strange particles ratio has been calculated and plotted in Fig.~\ref{fig:BaryonbypionMode0} (right) for different variants of the \texttt{PYTHIA8} model, such as the MPI-based CR, QCD-based CR and RH models and compared with the ALICE data \cite{PartRatio1}. The MPI-based CR underestimates the experimental data for strange hadrons except for K$^{0}_{s}$/$\pi$. The strange baryons ($\Lambda$, $\Xi$ and $\Omega$) to pion ratios estimated with the QCD-based CR model also underestimate the ALICE data except for $\Lambda$/$\pi$ and K$^{0}_{s}$/$\pi$ ratios. On the other hand, the model based on the mechanism of the rope formations reproduces the ALICE data quite well for all the studied particle ratios. However, there is a slight disagreement between the experimental data and \texttt{PYTHIA8} RH model at higher multiplicities for $\Lambda$/$\pi$ and $\Omega$/$\pi$ ratios. This discrepancy in $\Lambda$/$\pi$ and $\Omega$/$\pi$ ratios can be attributed to the higher production of $\Lambda$ baryons and lower yield of the pions at higher multiplicities in the \texttt{PYTHIA8} RH scenario. It is to be noted that in the RH scenario, the allocation of energy during the hadronization process, which includes the conversion of color flux tubes into hadrons, generally favors the production of heavier particles containing strange quarks over others. Since the energy available for hadronization is preferentially used for the creation of heavier particles, it may result in a lower pion yield. Similar results were also reported in the Ref.\cite{smita} where pion yields are found to be less in the rope hadronization scenario in \texttt{PYTHIA8} model. 

We have also estimated the ratio of yields of strange baryons with respect to strange meson (K$^{0}_{s}$) using the various models of \texttt{PYTHIA8}. These ratios are plotted as a function of charged-particle multiplicity and their comparison with the ALICE data is shown in Fig.~\ref{fig:BaryonbyK0s_Mode0}. One can observe from Fig.~\ref{fig:BaryonbyK0s_Mode0} (left) that the \texttt{PYTHIA8} model with MPI-based CR mechanism do not at all explain the experimental data on $\Lambda$/K$^{0}_{s}$, $\Xi$/K$^{0}_{s}$ and $\Omega$/K$^{0}_{s}$ ratios. The incorporation of the QCD-based CR mechanism though improved the agreement with the ALICE results, due to the production of more baryons, still could not able to explain the experimental data. One can note that the QCD-based CR model though can quantitatively describe the measured value of $\Lambda$/K$^{0}_{s}$ ratio \cite{multpp13} involving single strange quarks, it is very far from explaining the baryon-to-meson (K$^{0}_{s}$) ratios involving multi-strange quarks. Furthermore, the inclusion of color ropes in the  QCD-based CR very well reproduces the ratio of yield of multi-strange baryons ($\Xi$ and $\Omega$) to yield of K$^{0}_{s}$. This quantitative description by the RH model can be attributed to the enhanced production of strange quarks and di-quarks which leads to the enhanced production of multi-strange baryons. It is to be noted that the rope formation mechanism overestimates the $\Lambda$/K$^{0}_{s}$ ratio. This observation can be attributed  to the overestimation of $\Lambda$ baryon by the RH scheme of \texttt{PYTHIA8} as already shown in Fig.~\ref{fig:dndy_RR18}.

\section{Summary and Conclusion}\label{sec:5}
A comprehensive study has been carried out to have a better understanding of strange particle production in p+p collisions at LHC energy. The \texttt{PYTHIA8} model with different physics mechanisms has been employed to study the multiplicity-dependent $\langle dN/dy \rangle$, $\langle p_{\rm T} \rangle$, and the ratio of strange hadrons to mesons and compared with the available experimental data. The multiplicity-dependent $\langle dN/dy \rangle$ for K$^{0}_{s}$ can be nicely explained by the MPI-based color reconnection mechanism in \texttt{PYTHIA8}. However, it underestimates the production of strange baryons ($\Lambda$, $\Xi$, and $\Omega$). The disagreement with experimental data increases with increasing strange quark content. Furthermore, the reconnection range played no significant role as far as the yield of strange particles is concerned. However, the QCD-based CR model, which incorporates the junction topology, poorly describes the integrated yield of multi-strange baryons ($\Xi$ and $\Omega$) but can describe the $\langle dN/dy \rangle$ of strange hadrons ($K^{0}_{s}$ and $\Lambda$) containing single strange quarks. Moreover, the rope hadronization model very well explains the multi-strange baryons and K$^{0}_{s}$ production, barring the integrated yield of $\Lambda$ baryon at high multiplicity.

The mass-dependent hardening of the $p_{\rm T}$-spectra which are believed to be the signatures of collective behavior, as already reported for small systems at LHC energies, is observed in the current study performed on strange hadrons using the \texttt{PYTHIA8} model. One can observe that the RH coupled with the QCD-based CR scheme of \texttt{PYTHIA8} can not reproduce the multiplicity-dependent $\langle p_{\rm T} \rangle$. A slightly better description of experimental data has been provided by the MPI-based CR mechanism with $\mathrm{RR} = 1.8$. However, the  MPI-based CR mechanism with $\mathrm{RR} =  3.6$ shows a good agreement with the experimental data for strange hadrons. The magnitude of the RR in the CR mechanism plays a very important role in explaining the collective behavior of the produced particles and can mimic the effects as observed in heavy-ion collisions. 

The \texttt{PYTHIA8} model with the rope formation mechanism can give the best description of hadron-to-pion ratios as a function of $\langle dN_{ch}/d\eta \rangle$. However, it overestimates the $\Omega$/$\pi$ ratio at higher multiplicities which may be due to the production of less number of pions at higher multiplicities by the rope hadronization model. The \texttt{PYTHIA8} rope hadronization mechanism also mimics the multiplicity-dependent strange baryons to K$^{0}_{s}$ meson ratio as observed in ALICE. The ratio, on the other hand, estimated by other variants of the \texttt{PYTHIA8} model underestimates the experimental data. 

The following conclusions can be drawn from the present study:
\begin{itemize}
    \item MPI-based CR mechanism with $\mathrm{RR} = 3.6$ can describe the experimental data on multiplicity-dependent $\langle p_\mathrm{T} \rangle$ of strange hadrons.
  
    \item Rope hadronization scheme with QCD-based CR mechanism can explain the integrated yield of strange particles and bayon-to-meson ratios.
    
    \item  The \texttt{PYTHIA8} model, therefore, in general, cannot give a simultaneous and quantitative description of both the strangeness enhancement and collectivity for strange particles in p+p collisions. 
\end{itemize}

It is to be noted that, as reported in Ref.~\cite{Dhanjay}, the initial state phenomena such as MPI and the final state effects like the CR mechanism have significant roles to play at higher multiplicities and higher center-of-mass energies as far as the production of heavy flavor mesons such as $J/\psi$ is concerned. It is therefore interesting to see the effect of various CR modes in \texttt{PYTHIA8} on the production of heavy flavors at LHC energies.



\section*{Acknowledgments}
The authors would like to thank Christian Bierlich for providing the parameters for the rope hadronization model of \texttt{PYTHIA8}. The authors also acknowledge Prof. Sukalyan Chattopadhyay and Mr. Omvir Singh for their valuable suggestions.


\end{document}